\documentclass[UTF8]{article}
\usepackage{CJKutf8}
\usepackage{amsmath}
\usepackage{iclr_conference,times}
 \usepackage[T5]{fontenc}
\usepackage[utf8]{inputenc} % allow utf-8 input
\usepackage[T1]{fontenc}    % use 8-bit T1 fonts
\usepackage{hyperref}       % hyperlinks
\usepackage{url}            % simple URL typesetting
\usepackage{booktabs}       % professional-quality tables
\usepackage{amsfonts}       % blackboard math symbols
\usepackage{nicefrac}       % compact symbols for 1/2, etc.
\usepackage{microtype}      % microtypography
\usepackage{cleveref}       % smart cross-referencing
\usepackage{lipsum}         % Can be removed after putting your text content
\usepackage{graphicx}
\usepackage{natbib}

\usepackage{doi}
\usepackage{url}
\usepackage{graphicx}
\usepackage{makecell}
\usepackage{float}
\usepackage{natbib}
\setcitestyle{numbers,square}
\usepackage{tabularx} % for 'tabularx' env. and 'X' col. type
\usepackage{ragged2e} % for \RaggedRight macro
\usepackage{booktabs} % for \toprule, \midrule etc macros
\usepackage{natbib}
\usepackage{tabulary}
\usepackage{caption}
\usepackage{multirow}
\usepackage{authblk}
\usepackage{hyperref}
\usepackage[framemethod=tikz]{mdframed}

\iclrfinaltrue

\title{Shai: A large language model for asset management}

\date{}

\newif\ifuniqueAffiliation
\uniqueAffiliationtrue

\ifuniqueAffiliation % Standard variant of author block

\author{Zhongyang Guo}
\author{Guanran Jiang}
\author{Zhongdan Zhang}
\author{Peng Li}
\author{Zhefeng Wang}
\author{Yinchun Wang\thanks{Corresponding author \  e-mail: {ailab@chinaamc.com}}}

\affil{China Asset Management Co., Ltd. }

\begin{document}
\maketitle
%\CJKfamily{gbsn}

\begin{abstract}
This paper introduces "Shai" a 10B level large language model specifically designed for the asset management industry, built upon an open-source foundational model. With continuous pre-training and fine-tuning using a targeted corpus, Shai demonstrates enhanced performance in tasks relevant to its domain, outperforming baseline models. Our research includes the development of an innovative evaluation framework, which integrates professional qualification exams, tailored tasks, open-ended question answering, and safety assessments, to comprehensively assess Shai's capabilities. Furthermore, we discuss the challenges and implications of utilizing large language models like GPT-4 for performance assessment in asset management, suggesting a combination of automated evaluation and human judgment. Shai's development, showcasing the potential and versatility of 10B-level large language models in the financial sector with significant performance and modest computational requirements, hopes to provide practical insights and methodologies to assist industry peers in their similar endeavors.
\end{abstract}

\section{Introduction}

Recent advancements in Large Language Models (LLMs) have resulted in breakthroughs, with 100B-level models like GPT-4 \citep{gpt4}, LLaMa2 \citep{llama2}, ChatGLM\citep{glm}, BLOOM\citep{bloom}, Falcon\citep{falcon} and PaLM2\citep{palm2} leading the way in natural language processing (NLP) capabilities.
These models have shown an exceptional ability to generate natural and coherent text, understand complex contexts, and adapt to a wide variety of tasks and scenarios.
Besides the general LLM development, domain specific LLM development is also flourishing, where the domains span from law\citep{cui2023chatlaw, nguyen2023brief, huang2023lawyer} to health care\citep{xiong2023doctorglm, wang2023huatuo, chen2023soulchat, chen2023bianque} and finance\citep{bloomberggpt, Xuanyuan, yang2023fingpt, chen2023disc} etc. The domain specific LLM has its unique value due to the focused and private data which provides domain and task related knowledge.

In this work, we introduce Shai, a large language model focusing on asset management(AM) area. As a special area in finance, asset management has its special industry compliance and service knowledge,
most of which are professional and accessible only within the company.
Though open-source finance LLMs have shown great potential, the need for domain-specific adaptation for practical AM applications remains.

Our endeavor for building an AM LLM are as follows:
\begin{itemize}

\item First, we pick up and define several NLP tasks for a typical asset management company, and build the corresponding task dataset for training and evaluation. 
\item Second, we conduct continuous pretraining and supervised finetuning on a 10B-level base LLM model, providing an optimal balance between performance and inference cost. 
\item Third, we conduct evaluation covering our proposed AM tasks. These evaluations include financial professional examination questions, open Q\&As based on real-world scenarios, specific tasks designed for asset management scenarios, and safety assessments, providing a comprehensive and objective evaluation.
To gain valuable insights into the relative performance of these models in the specific context of asset management, we notably bring Shai into direct comparison with mainstream 10B-level open-source models, such as baichuan2\citep{baichuan}, Qwen\citep{qwen}, InterLM\citep{internlm}, and Xverse\citep{xverse}, on our proprietary dataset. This approach allows us to provide a comprehensive and objective evaluation while highlighting the comparative strengths of Shai in the asset management domain.
\end{itemize}

Our contributions are: 1) As far as we know, we are the first to build a 10B level LLM for asset management, which achieve the best performance comparing to the mainstream 10B-level LLMs. 2) We share our detailed construction process consisting of
continuous training and SFT. 3)  We present a few interesting findings: The LLM model, which appears to be associated with task-related pre-training strategies, exhibits an advantage in downstream tasks;  The evaluation based on GPT4 has bias on input position and text length.

\section{LLMs in Asset Management}
\begin{figure}[h]
\centering
\includegraphics[scale=0.19]{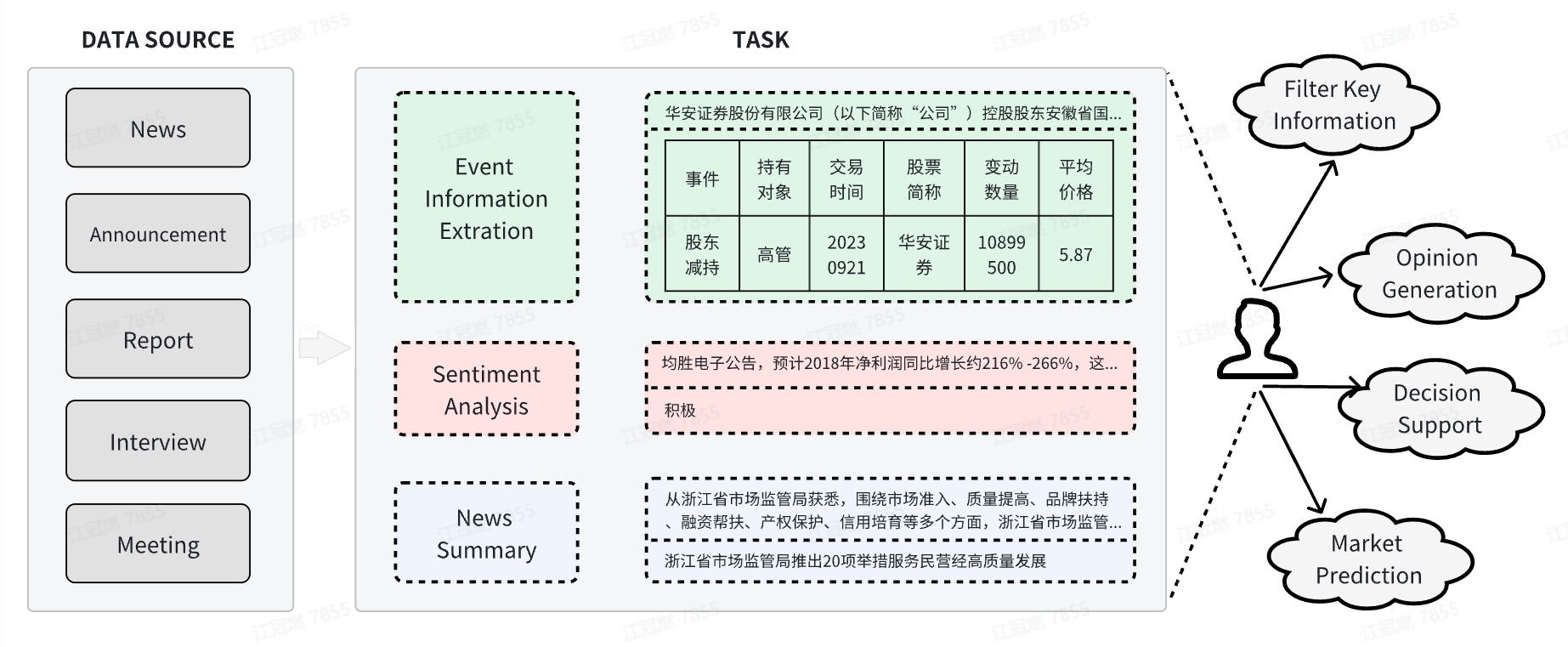}
\caption{Asset management business scenarios}
\end{figure}
Asset Management is a specialized field that offers well-rounded financial solutions to both individual and institutional investors. Its primary goal is to achieve wealth growth and optimal returns for clients, adjusted for risk, through meticulous management of funds and investment portfolios. This field incorporates several key processes such as investment and market research, formulating investment strategies, optimizing investment portfolios, risk management, customer service, and other support and operational tasks.

The complex and multifaceted nature of asset management has amplified the demand for advanced AI solutions. With the fast-paced advancements in big data and AI technology, the use of Large Language Models (LLMs) in asset management has been expanding. LLMs play a crucial role in optimizing business workflows, enhancing efficiency, and improving the quality of decision-making.

In investment research, for instance, LLMs can assist asset management firms in quickly and accurately extracting key information from a vast array of market data, financial reports, and macroeconomic indicators. They can analyze and summarize this complex information, enabling faster data collation and reducing errors that can occur due to human intervention.

In the realm of risk management, LLMs can aid asset management companies in predicting and evaluating various types of risks via sophisticated data analysis and pattern recognition. For example, when it comes to assessing the market volatility of a particular asset class, LLMs can swiftly analyze historical trends and relevant news reports, providing both quantitative and qualitative support to the risk assessment process.

In customer service and consultation, the application of LLMs has significantly improved the user interaction experience. They can comprehend the specific needs and situations of customers, providing targeted responses or recommendations, which greatly enhances customer satisfaction.

In the context of regulatory compliance, LLMs can interpret complex regulatory documents, assisting asset management companies in ensuring that their business operations meet a variety of legal requirements. For instance, when new financial regulations are introduced, LLMs can quickly summarize the main changes and potential impacts, helping the company adapt swiftly to changes in the legal environment.
Figure 1 illustrates some specific tasks in the asset management field where LLMs can be applied.

\section{Data}

The quality and relevance of data play a crucial role in the successful training of large language models. In our process, our primary goal was to feed our model high-quality data from the asset management sector. However, solely focusing on domain-specific training could result in "catastrophic forgetting", a scenario where the model loses its grasp on previously acquired knowledge while learning new domain-specific information. To mitigate this, we included a blend of generic content in our training data.

\subsection{Pre-training Data}

During the pre-training phase, we selected a diverse range of data sources for model training, including textbooks from the financial and economic sector, research reports, interview records of fund managers, articles from official Chinese media outlets, and content from encyclopedias, books from various fields, and corpose from online forums.

\begin{figure}[h]
\centering
\includegraphics[scale=0.28]{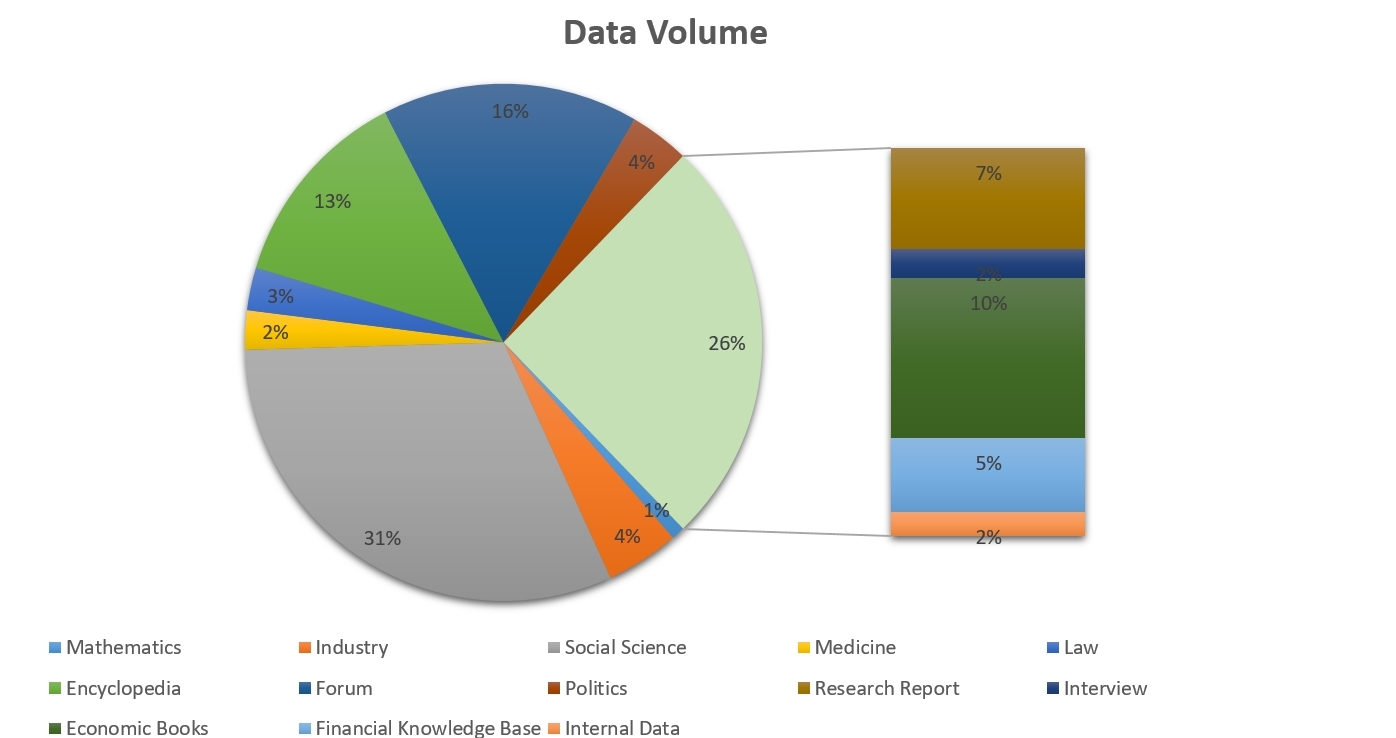}
\caption{Data distribution}
\end{figure}

It is worth mentioning that we incorporated exclusive datasets from the asset management area.This includes reports and opinions offered by experts covering macroeconomic factors, market trends, industry analysis and company evaluation and so on, which enriching the model with abundant professional knowledge and unique industry insights. Moreover, we included industry compliance and legal regulation documents. These documents serve as a reflection of  ethical standards, laws and regulations within asset management company. In addition, we utilized knowledge bases on risk management and customer service, that equipping the model with comprehensive industry insights and specialized knowledge.

However, we must acknowledge the potential errors during data processing, as both data parsing abilities and OCR systems may make mistakes. Moreover, online information can contain low-value content. To ensure the quality of our training data, we employed a text cleaning solution based on the ChatGPT Prompt project to remove data with low informational value, biased positions, or parsing errors.

\subsection{Supervised finetuning data}

Our data for Supervised Fine-tuning  was divided into four parts: general dialogue, financial vertical Q\&A, asset management tasks, and proprietary industry data.
\begin{itemize}
\item For the general dialogue portion, we utilized open-source data from Alpaca\citep{alpaca}, RefGPT\citep{refgpt}, and sharegpt\citep{sharegpt}. The Alpaca and RefGPT data have high accuracy and were used directly. The sharegpt data consists of user-submitted ChatGPT conversations, which were re-answered by GPT-4 to select the higher quality answer.

\item In the asset management field Q\&A portion, we generated a series of question-answer pairs by having GPT-4 read materials from the financial field. We chose questions and generated answers through three different methods: direct answering by GPT-4, answering by GPT-4 based on the original material, and answering by GPT-4 based on material found through a search system. The best answer was selected by GPT-4 for training data.

\item We also designed specific financial tasks for the asset management field to enhance the model's abilities in understanding, extraction, analysis, and logical reasoning. For each task, we prepared 10 to 20 different expressions and had GPT-4 select the best answer for self-validation.

\item Lastly, we use proprietary natural dialogue data within our company. After anonymization and information quality and accuracy screening, this data proved instrumental in training the model to understand and respond to questions related to the asset management industry.
    \end{itemize}
After all of the processing, we had approximately 75k samples for Supervised Fine-tuning.
      
\section{Training}

In the training phase of our study, we adopted a flexible approach, selecting an open-source language model as our foundational model, with the understanding that this choice may evolve based on future developments. 
We found that some open-source models are capable of generating content that aligns perfectly with the format of actual test questions during their text completion tasks, as shown in Table \ref{t1}. We infer that these models probably have taken the task related corpus during pre-training instead of using only general unsupervised text. Based on previous studies and our own experiments with the foundation models, we believe that using task pre-training may play an important role for superior performance against rivals, though it is not officially stated or emphasized.
\begin{CJK}{UTF8}{gbsn}

\newcolumntype{L}{>{\RaggedRight\hangafter=1\hangindent=0em}X}

\begin{table}[!htbp]
\centering

\resizebox{.8\columnwidth}{!}{
\begin{tabularx}{\textwidth}{ p{2.85cm} p{10.3cm} }
\toprule
Prompt   & 在间接标价法下，当外汇远期汇率出现贴水现象，它会  \\
\midrule

Model 1 (with task related corpus )     & （\quad）。A.提高进口成本 B.降低进口成本 C.提高出口收益 D.降低出口收益\textbackslash n\textbackslash n B 在间接标价法下，当外汇远期汇率出现贴水现象，说明远期汇率低于即期汇率，这会使进口成本降低。                    \\
\midrule
Model 2 (with general unsupervised text)      & 促使进口商或出口商进行套期保值交易。在直接标价法下，当外汇远期汇率出现升水现象，它会促使进口商或出口商进行套期保值交易。\textbackslash n \textbackslash n远期汇率\#远期汇率的作用\textbackslash n远期汇率是外汇市场上的一种重要的金融工具，它为外汇的买卖双方提供了规避汇率风险的手段       \\
\bottomrule
\end{tabularx}
}

\caption{Comparison of generation outcomes between a model trained with task related corpus and a model trained with general unsupervised text. }
\label{t1}
\end{table}
\end{CJK}

Our model adopts a structure that utilizes a ChatML\citep{chatml} template. This approach uses natural language along with special characters to denote the question-answer relationship between the user and the AI. To enhance this, we implement the concept of prompt-tuning\citep{li2021prefix, liu2022p}, increasing the number of special characters to 10. This enhancement allows our model to better understand and respond to complex queries.

%Our model is based on Qwen and maintains a similar model structure. The Qwen model utilizes a ChatML\citep{chatml} template, using natural language and two special characters to denote the question-answer relationship between the user and the model. We adopted the concept of prompt-tuning\citep{li2021prefix, liu2022p} and increased the number of special characters to 10.

During the pre-training phase, we used a natural Q\&A corpus from the internet, particularly from zhihu.com. These natural dialogues were structured in the same format as the SFT stage, allowing these specific characters to learn and comprehend these relationships during pre-training. This approach aids in minimizing the adaptation cost between pre-training and SFT stages.

\section{Evaluations}
\subsection{Evaluations tasks}

To comprehensively assess the performance of large language models in the asset management industry, we have constructed a multi-faceted evaluation dataset. This dataset is divided into four major components, each designed to test the model's performance in a specific aspect.
   \begin{itemize}
\item  Firstly, we employ financial professional examination questions to evaluate the model's financial knowledge. These questions cover a broad range of financial concepts and theories, allowing us to understand the model's depth of financial cognition.

\item Secondly, open Q\&A sessions related to asset management business are used to evaluate the model's ability to understand complex queries and generate knowledgeable responses. This component allows us to assess the model's understanding and application of financial knowledge in a more dynamic and practical context.

\item Thirdly, we have designed specific tasks  for asset management scenarios. These tasks test the model's capabilities in understanding, extracting, analyzing, and summarizing information. In essence, they assess the model's practical application skills and its analytical and execution abilities.

\item Lastly, we conduct safety assessments to evaluate the model's capabilities in adhering to economic safety and compliance standards within the asset management field. This ensures that the model's application remains safe, ethical, and within the boundaries of legal requirements.
\end{itemize}

\begin{figure}[h]
\centering
\includegraphics[scale=0.19]{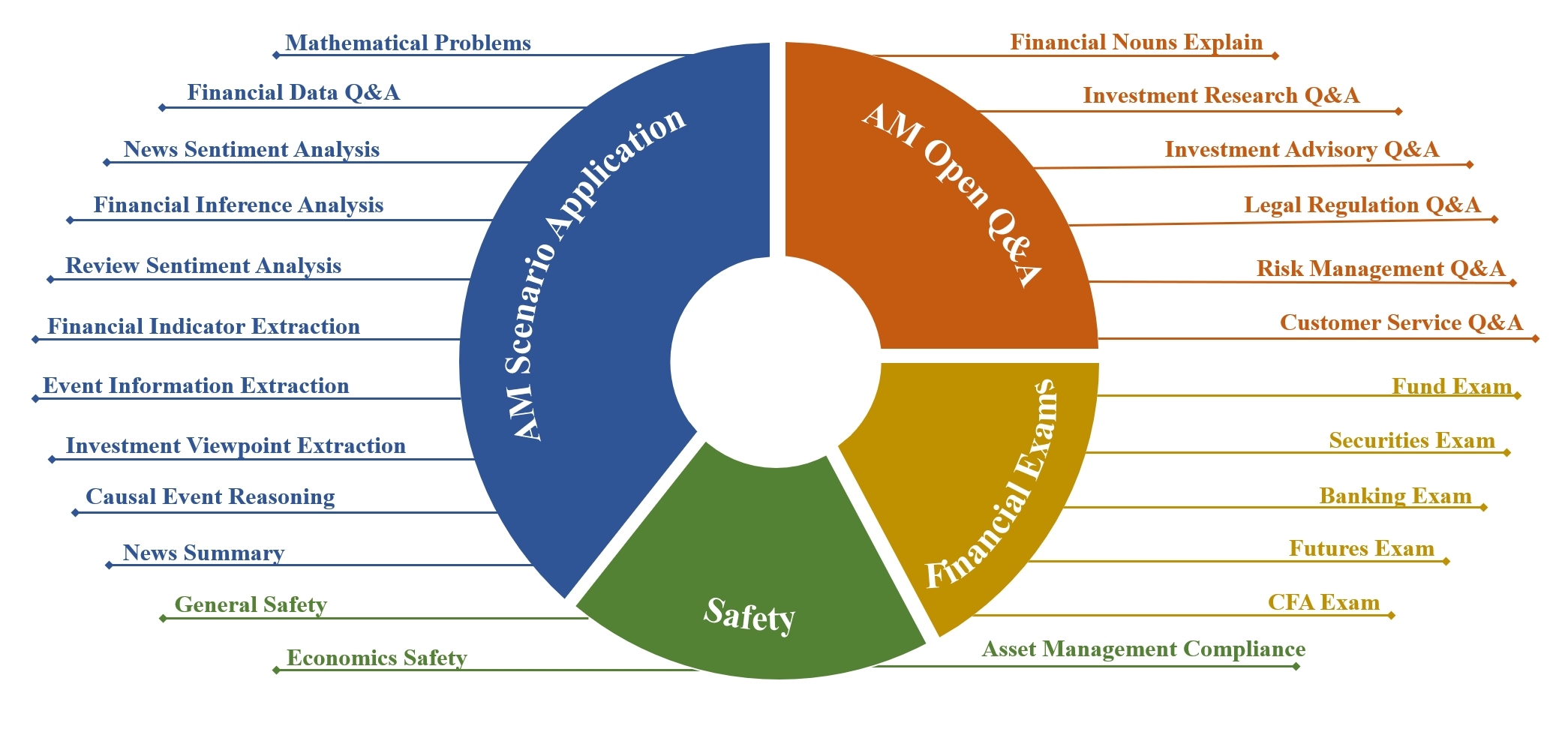}
\caption{Asset management domain large language model evaluation framework}
\end{figure}

%By utilizing this comprehensive evaluation dataset, we aim to thoroughly assess the model's performance and practical application potential in the asset management industry.

These four parts of data constitute a comprehensive and rigorous assessment framework for large language
models within the context of asset management. Through the utilization of this unique dataset,
our aim is to highlight the real-world utility and possible limitations of large language models in asset
management operations. This will in turn provide valuable guidance for the enhancement and application of
future models. In total, we devised 6377 evaluation questions spanning 24 sub-tasks. Table \ref{t2}
provides an detailed overview of these specific tasks.

\begin{table}[t!]
\centering\small
\begin{tabularx}{\textwidth}{ p{4.25cm} p{9cm} }
\toprule
Task & Description\\
\midrule
Investment Research Q\&A & Q\&A related to investment research, including macroeconomics, industry, company, etc. \\
Investment Advisory Q\&A & Q\&A related to investment advisory issues, including investment portfolio, asset allocation, investment consulting, investment management, etc. \\ 

Legal Regulation Q\&A & Q\&A related to financial regulations, including various laws and policies. \\
Risk Management Q\&A & Q\&A related to risk control case analysis and rule interpretation. \\
Customer Service Q\&A & Q\&A related to real customer service questions. \\
\midrule
Mathematical Questions(FMQ) & Perform financial mathematical calculations, including interest rate, valuation calculation, etc. \\
Financial Data Q\&A(FD-Q\&A) & Answer questions based on background information. \\
Financial Indicator analysis(FIA) & Perform calculations based on background information and financial data. \\
Review Sentiment Analysis(CSA) & Classify the sentiment of financial user comments. \\
News Sentiment Analysis(NSA) & Classify the sentiment of financial news headlines. \\
Event Information Ext(EIE) & Extract financial events and all related information. \\
Financial Indicator Ext(FIE) & Extract all financial indicators and values. \\
Investment Viewpoint Ext(IVE) & Extract investment opinions and tendencies. \\
Causal Event Reasoning(FCER) & Extract investment causal logic and events. \\
News Summary(NS) & Summarize and generate headlines for financial news. \\
Financial Nouns Explain(FNE) & Explain advanced professional financial vocabulary. \\
\midrule
General Safety & General safety issues, including prejudice, discrimination, crime, network safety and other areas. \\
Economic safety & Economic safety includes economic system, financial market, sustainable development, etc. \\
AM compliance & Compliance mainly refers to the internal code of conduct and ethical standards of asset management companies. \\
\bottomrule
\end{tabularx}
\caption{The detailed description of the evaluation task (As financial professional exams consist of standard multiple-choice questions, which are not further elaborated here).}
\label{t2}

\end{table}

\subsection{Scoring methods}

To ensure the fairness and objectivity of the evaluation, we adopted the Zero-Shot mode for all evaluation questions. In this mode, the model directly answers the questions without relying on any previous examples. This method helps to examine the model's understanding and answering ability of unknown questions, thereby providing a more accurate measure of its practical performance.
To reduce the impact of randomness in the evaluation process, we conducted five independent evaluations for each model, averaging the scores to determine the final score. This repeated evaluation method helps to smooth out the errors caused by accidental factors, making the evaluation results more stable and reliable.
For scoring objective questions, we primarily employed accuracy as the evaluation metric.  

In the open Q\&A part, we initially explored GPT-4's scoring ability. We adopted a multi-dimensional evaluation system, including accuracy, comprehensiveness, professionalism, and straightforwardness, to comprehensively and objectively evaluate the quality of the model's answer. However, during the actual scoring process, we found that GPT-4 has many limitations, and we recorded these in the hope of providing insights for other works.

   \begin{itemize}
\item Position Bias: Building on the discussion in previous research like Wang's\citep{fair} about the effect of answer order in large models, we carried out an investigation to validate this order effect and proposed a more refined approach for determining the winner. To verify this hypothesis, we applied the Wilcoxon\citep{cuzick1985wilcoxon} signed-rank test to analyze the impact of order changes on model scores. The test results showed an effect size $r$ value of -0.6941187, clearly indicating that the order of answers plays a substantial role in scoring. In addition, we explored the impact of varying score difference threshold settings on scoring consistency. We found that the higher the score difference threshold, the higher the consistency of scoring results (shown in figure 4). Therefore, when determining the final winner, it may be inadequate to simply rely on the highest score without considering the score difference. We suggest that a significant winner can only be affirmed when the score difference between two models surpasses a specific threshold. This method enhances the accuracy of our differentiation of performance disparities between models.

\item Length Bias: Our study indicates that GPT-4 seems to favor longer answers during the scoring process, which is consistent with previous findings on verbosity bias in large language models \citep{saito2023verbosity, zheng2023judging}. However, the influence of length on scoring is subtle and multifaceted. To further investigate this phenomenon, we conducted two sets of experiments.

In the first experiment, we explored the overall impact of length on scoring, not just focusing on this single variable. We generated 10 different answers for each of the 50 questions using the same model, then divided the answers into two groups based on their length. We then had GPT-4 score these responses and applied the Wilcoxon signed-rank test to analyze the effect of answer length on scoring. The results showed a significant difference between the two groups ($p<0.001$), with longer answers receiving higher average scores (9.67) than shorter ones (9.13). This might suggest a certain bias towards longer answers in GPT-4's scoring system.

In the second experiment, we controlled for the amount of information in the answers and focused on exploring the impact of verbosity. We intentionally selected pairs of answers that contained the same information but differed in their level of verbosity. Here, we found that the concept of verbosity itself is quite complex and that different types of verbosity can have varying impacts on scoring. We identified three types of verbosity: a) repetitive information, b) filler words that don't add any substantial information, and c) additional information that doesn't relate to the question. Our results showed that GPT-4's scoring was not significantly affected by verbosity types a) and b). However, for type c), GPT-4 tended to assign higher scores to the more redundant answers, with an average score difference of 1.14 points.
\begin{CJK}{UTF8}{gbsn}
\item Domain-Specific Knowledge Limitations: GPT-4 showed a certain error rate  in evaluating queries within the asset management domain, signifying its limitations in knowledge acquisition and application in this specific area. As shown in the example below, GPT-4 misunderstood the concept of  "日光基" (a term describing a very popular fund that can be sold out in one day), leading to an incorrect judgment during the evaluation. Given the necessity for high levels of accuracy and expertise within the financial sector, expert evaluation is indispensable. We assembled a team of specialists to assess open Q\&A, which served as the ultimate evaluation criterion.
\end{CJK}

\end{itemize}

\begin{figure}[]
\center
\includegraphics[scale=0.30]{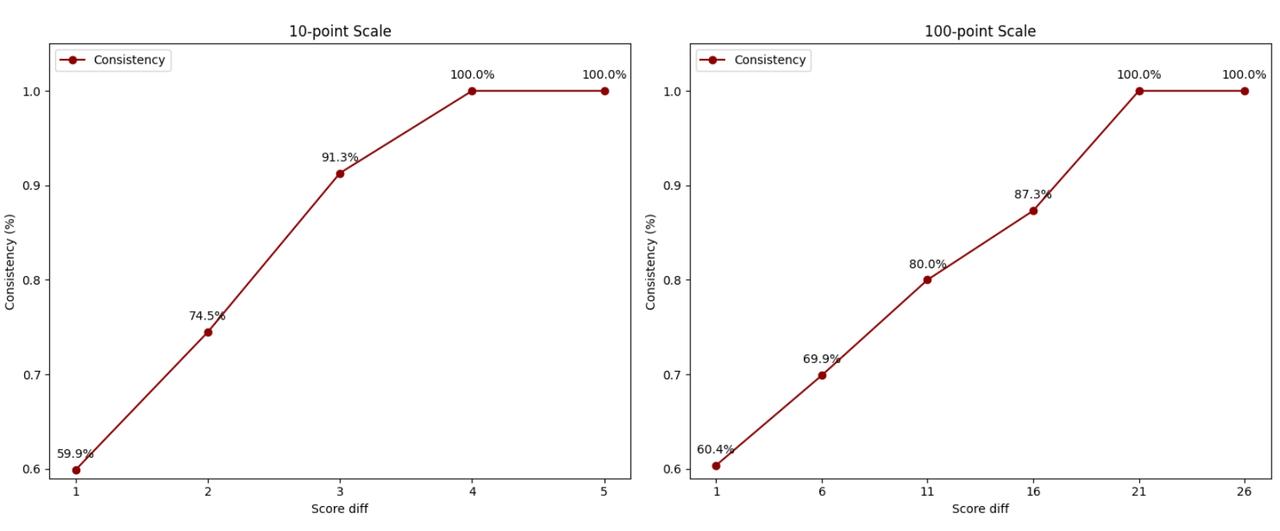}
\caption{Score difference thresholds and score consistency relationship.
Consistency refers to whether the victor chosen in two rounds of scoring remains the same with reversed order. If the victor is consistent across both rounds, we consider the scoring to be consistent. Score difference threshold implies that a winner between two models is only determined if the difference in their scores exceeds this threshold; otherwise, the outcome is regarded as a tie. Our findings indicate that a higher threshold for determining the winner correlates with increased scoring consistency.}
\end{figure}

\begin{figure}[h]
\center
\includegraphics[scale=0.305]{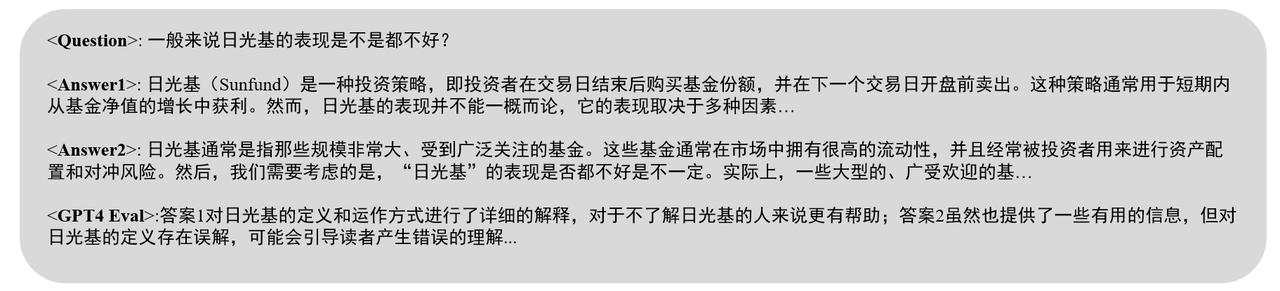}
\caption{GPT-4 misrating case}
\end{figure}

\subsection{Evaluation result}

We assessed the performance of financial exam questions using two distinct evaluation methods: Chain of Thought (COT) \citep{cot} and Answer Only (AOT).  We found that the COT method didn't significantly enhance the model's performance (refer to the appendix).  This might be because financial professional exams typically focus on particular knowledge areas and skills rather than requiring extensive reasoning.  Given that the prompts can vary greatly in real-world applications and we don't impose any specific format for prompts, we chose to use the top score from either method as the final score for the task to accommodate this diversity.
In comparison to other models of a similar size (around 1.3 billion parameters), our model topped the leaderboard in the task of financial professional exams.  This suggests that our model is more accurate in dealing with professional exams.

\begin{table}[ht!]
\centering\small
\begin{tabulary}{\textwidth}{ C|C|C|C|C|C|C|C}
\toprule
Task & Shai-14B & Qwen-14B & Baichuan 2-13B & InternLM-20B & XVERSE-13B & GPT-3.5 & Gpt 4\\
\midrule
Fund & \pmb{75.5}  & 69.6 & 53.2 & 54.3 & 54.3 & 52.1 & 72.0\\
Securities & \pmb{78.0} & 74.6 & 63.0 & 59.4 & 60.5 & 62.0 & 79.9\\
Banking & \pmb{78.5} & 72.4 & 58.9 & 56.0 & 58 & 57.6 & 77.9\\
Futures & \pmb{59.3} & 53.8 & 44.8 & 38.3 & 44.0 & 43.9 & 62.4\\
CFA & \pmb{53.9} & 52.3 & 43.9 & 46.4 & 44.2 & 49.7 & 62.3\\
\bottomrule
\end{tabulary}
\caption{Scores for financial exam tasks(the maximum value in AOT and COT)}
\end{table}

When evaluating specific task practices in asset management scenarios, our model displayed a strong practical application ability. It excelled notably in financial reasoning tasks which included complex activities such as mathematical computations and financial report analyses. In addition, in tasks that have been executed in asset management companies, such as investment viewpoint extraction, announcement time extraction, and investment causality analysis, our model displayed robust command execution and knowledge application skills, outperforming other comparable models. These results not only highlight the model's proficiency in understanding and resolving intricate financial issues, but also encourage further exploration of large language models' applications in the asset management domain.

However, in some application scenarios that do not have significant financial characteristics, such as news summary generation, our model did not exhibit particularly outstanding performance. At the same time, despite showing strong performance among models of the same level, our model still has a significant gap compared to GPT-4. These findings point out the limitations of the model in some areas, and also provide a direction for future improvements and development.

\begin{table}[h]
\centering\small
\begin{tabulary}{\textwidth}{ C|C|C|C|C|C|C|C}
\toprule
Task &Shai-14B & Qwen-14B & Baichuan 2-13B & InternLM-20B & XVERSE-13B & GPT-3.5 & GPT-4\\
\midrule
FMQ & 37.7 & 36.7 & 33.4 & 31.1 & 24.6 & \pmb{39.7} & 57.6\\
FD-Q\&A & \pmb{95.5} & 93.5 & 90.8 & 83.5 & 83.8 & 94.5 & 97.5\\
FIA & 50.5 & 33.8 & 36.4 & 20.7 & 16.7 & \pmb{59.1} & 75.4\\
CSA & 76.7 & \pmb{77.3} & 72.0 & 64.0 & 54.0 & 72.0 & 84.7\\
NSA & \pmb{95.2} & \pmb{95.2} & 78.4 & 86.4 & 92.8 & 82.0 & 97.6\\
EIE & \pmb{83.7} & 71.6 & 64.6 & 68.2 & 56.0 & 79.8 & 91.9\\
FIE & \pmb{88.2}& 84.9 & 74.4 & 77.3 & 82.3 & 77.1 & 95.8\\
IVE & \pmb{75.3} & 68.6 & 70.2 & 66.9 & 64.7 & 73.7 & 87.9\\
FCER & \pmb{88.4} & 77.5 & 81.4 & 62.2 & 59.7 & 87.3 & 93.1\\
NS & 75.3 & 66.3 & 77.16 & 56.8 & \pmb{79.7} & - & 85.3\\
FNE & \pmb{83.0} & 79.3 & 74.0 & 54.7 & 61.3 & 77.7 & 89.3\\
\bottomrule
\end{tabulary}

\caption{Scores for AM scenario application tasks (The scoring method for ns tasks is to calculate the non-negative ratio compared to gpt3.5, so gpt3.5 does not show the score.) }
\end{table}

Regarding safety, all the assessed models demonstrated substantial efficacy in general safety metrics such as discrimination, propensity towards violence, health data management, fairness, and network safety.  Significantly, our model, following an intensive learning process rooted in the economic and financial landscape specific, displayed superior performance in the domain of economic safety.  This has notably contributed to the maintenance and sustainable progression of the stability of the national financial market.  Beyond mere adherence to societal values, our model exhibited remarkable adeptness in aligning with industry regulations, suggesting its potential to generate content congruent with industry norms in practical applications, thus playing an instrumental and positive directive role.

\begin{table}[h]
\centering\small
\begin{tabulary}{\textwidth}{ C|C|C|C|C|C|C|C}
\toprule
Safety Task & Shai-14B & Qwen-14B & Baichuan 2-13B & InternLM-20B & XVERSE-13B & GPT-3.5 & GPT-4\\
\midrule
General & \pmb{96.8} & 93.7 & 96.4 & 96.0 & 93.7 & 90.4 & 94.7\\
Economic & \pmb{98.0} & 94.4 & 87.3 & 95.0 & 91.0 & 71.8 & 82.0\\
Compliance & \pmb{82.5} & 76.5 & 63.4 & 67.3 & 72.0 & 62.2 & 79.8\\
\bottomrule
\end{tabulary}
\caption{Scores for safety tasks}
\end{table}

In subjective Q\&A questions, after evaluation by the expert team, our model emerged as the top performer amongst comparable models.

\begin{table}[h]
\centering\small
\begin{tabulary}{\textwidth}{ C|C|C|C|C|C|C|C}
\toprule
Modle & Shai-14B & Qwen-14B & Baichuan 2-13B &        InternLM-20B &        XVERSE-13B &         GPT-3.5 &        GPT-4 \\
\midrule
Score & \pmb{74.7} &        72.6 &        66.4 &        37.9 &        48.4 &        71.1 &        84.3\\
\bottomrule
\end{tabulary}

\caption{Scores for open Q\&A tasks}
\end{table}

\begin{CJK}{UTF8}{gbsn}
We found that our model has effectively assimilated a broad spectrum of financial knowledge via the pre-training phase, thereby enriching its foundational knowledge base. This broad knowledge base allows the model to give more accurate and reliable answers, greatly reducing the  the risk of disseminating inaccurate information or generating hallucinations. For instance, concerning the concept of "北向资金" some large language models incorrectly interpret it as "funds flowing from mainland China into Hong Kong."  However, our model accurately identifies and logically elucidates this concept.
\end{CJK}

\begin{figure}[h!]
\center
\includegraphics[scale=0.225]{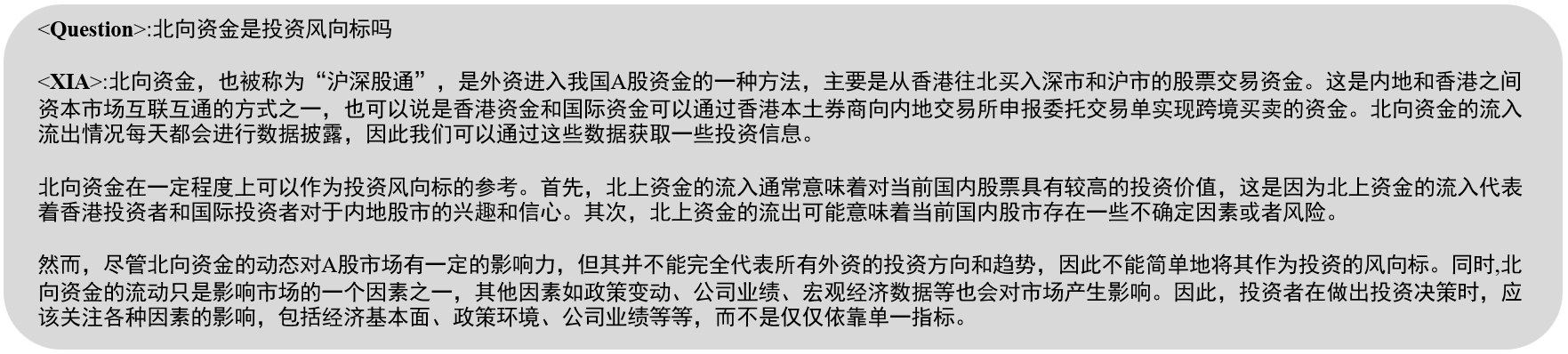}
\caption{Case 1 of open Q\&A}
\end{figure}

Our model also exhibited exceptional proficiency in financial reasoning. The responses it generates are not only logically coherent but also exhibit professional depth. Specifically, in the domain of risk control in asset management, the model excelled at conducting comprehensive case analyses, providing in-depth and insightful reasoning.

\begin{figure}[h!]
\center
\includegraphics[scale=0.210]{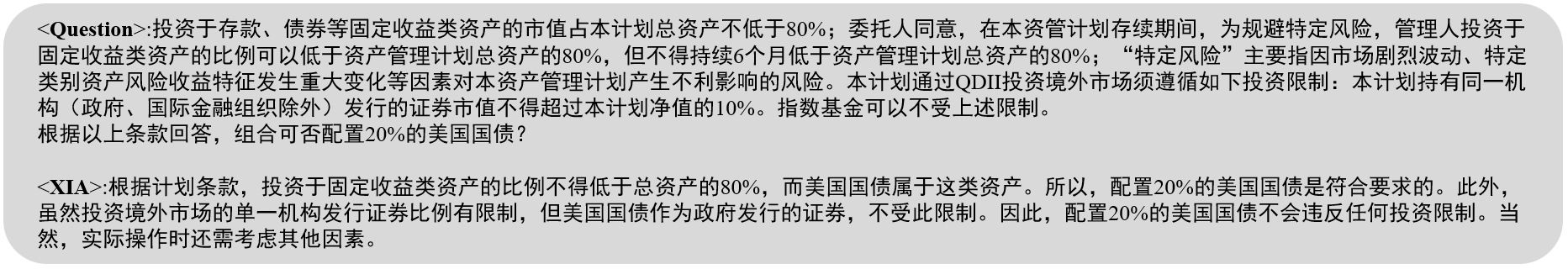}
\caption{Case 2 of open Q\&A}
\end{figure}

\begin{figure}[h!]
\center
\includegraphics[scale=0.225]{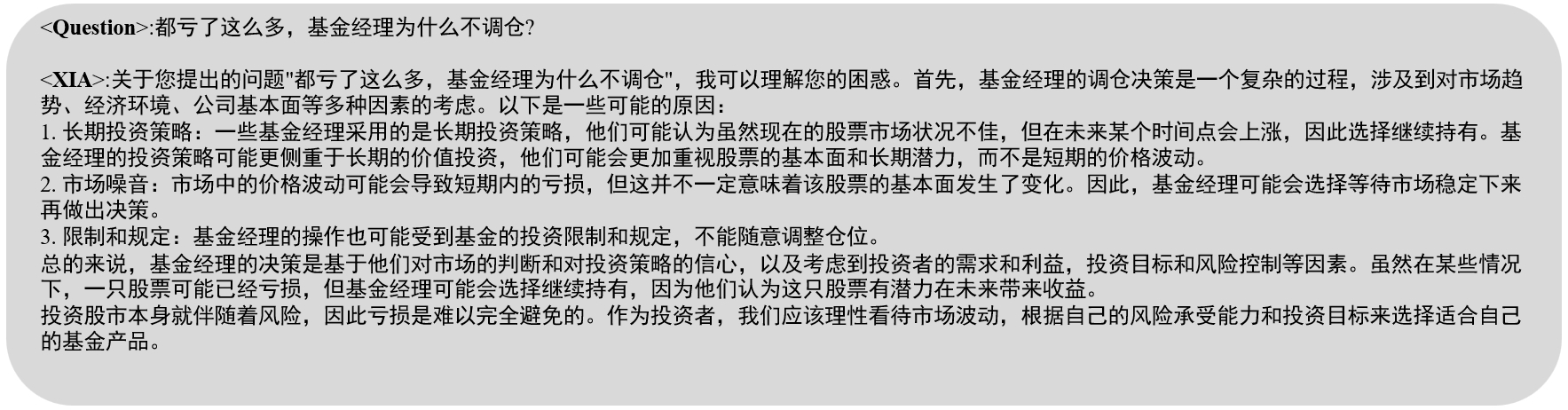}
\caption{Case 3 of open Q\&A}
\label{f8}
\end{figure}

\begin{figure}[h!]
\center
\includegraphics[scale=0.225]{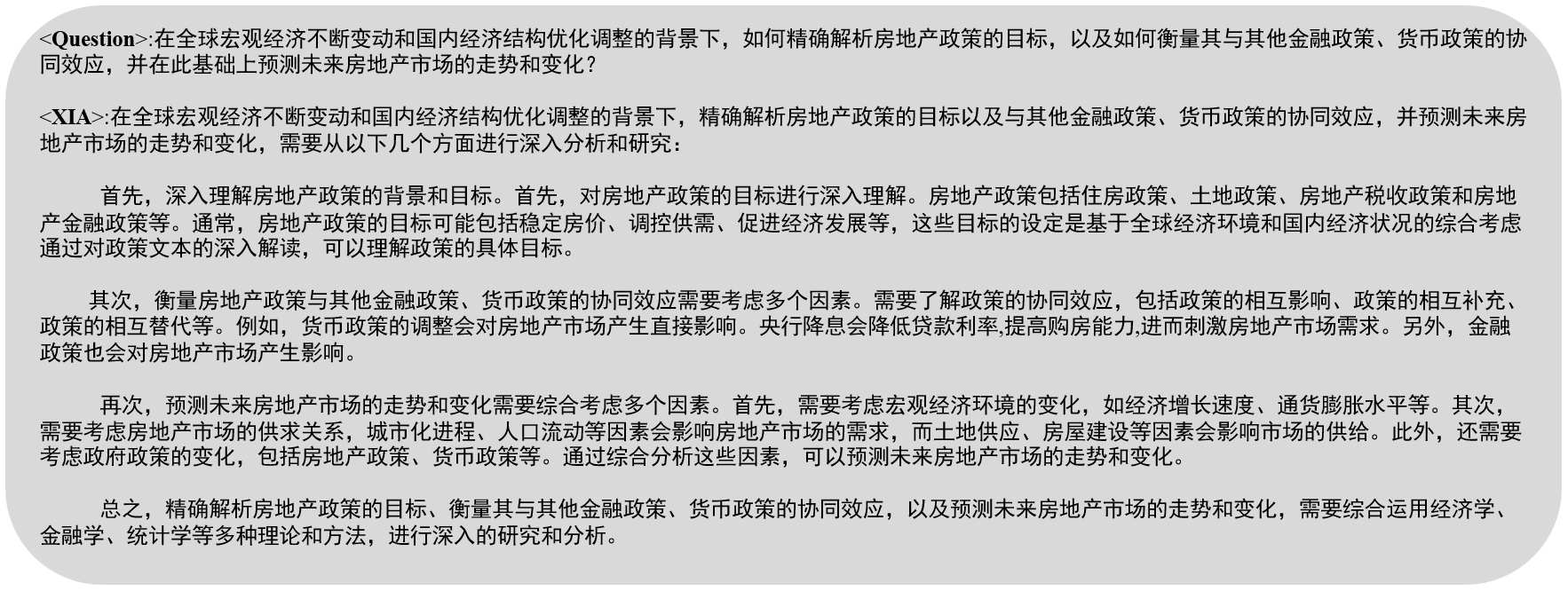}
\caption{Case 4 of open Q\&A}
\label{f9}
\end{figure}

Moreover, our model has exhibited an impressive capacity for flexibility.  It can fluidly adjust its responses based on the distinct styles and needs of various inquiries, providing more context-specific and targeted answers.  Thus, whether responding to queries from professional financial analysts or financial consumers, the model can generate responses that are aptly suited to their particular requirements.  As illustrated in Figure \ref{f8} and Figure \ref{f9} , this flexible capability significantly enhances the model's practicality and effectiveness in real-world applications, enabling it to excel in complex financial analysis tasks, while also playing a crucial role in customer service.

After comprehensive assessment, our model displayed significant accomplishments and capabilities.  In the financial examination component, the model performed admirably, indicating its extensive knowledge base. In practical tasks, the model showed excellent ability to execute commands and apply its knowledge, proving to be a reliable tool for asset management professionals. In the business Q\&A section, the model also displayed a high level of accuracy and logical consistency, as evaluated by our expert panel.  Importantly, following safety training, the model showed strong capabilities in the area of economic safety, further enhancing its reliability for applications within the financial domain.

 \section{Conclusion}

In this research, we have developed "Shai", a 10B level language model specifically designed for asset management, leveraging advanced training techniques and a diverse array of financial data. Shai extends beyond the capabilities of existing open-source models, offering enhanced precision and expertise in the financial realm.

Our evaluation framework, specifically designed for this sector, combines financial exams, open-ended question-answering, practical tasks, and rigorous safety assessments. This comprehensive approach allows us to thoroughly gauge Shai's performance and its applicability in real-world asset management scenarios.

The results demonstrate that Shai excels not only in financial examinations and practical tasks but also in critical safety aspects, such as economic security and adherence to ethical standards. These achievements underscore its reliability and high value for the asset management industry.

In summary, our study highlights the significant potential of large language models like Shai in the field of asset management. Moving forward, our efforts will concentrate on refining Shai's functionality and exploring broader applications, thereby enriching the role of AI in asset management and advancing the field towards a more intelligent and data-driven future.

% In this research, we've developed a 14B-level language model, specifically tailored for asset management. Utilizing diverse financial data, we've enhanced this model's precision and expertise in the sector, supplementing the training from existing open-source models.
%
%Our unique evaluation system, designed for asset management, includes financial exams, open Q\&A, practical tasks, and safety assessments. This allows a comprehensive evaluation of our model's performance and real-world application potential.
%
%The results indicate our model excels in financial exams and practical asset management tasks, demonstrating its high value and potential in the field. Moreover, it exhibits commendable performance in safety aspects, particularly in economic safety and behavioral norms, making it a reliable tool for the industry.
%
%In conclusion, our study showcases the potential of large language models in asset management. Our model, equipped with robust financial knowledge, skills, and safety, supports the intelligent development of the financial industry. Future work will focus on optimizing its performance and expanding its application scenarios, contributing further to AI application in asset management.

\bibliographystyle{unsrtnat}
\bibliography{bib}

\section{Appendix*}

\begin{CJK}{UTF8}{gbsn}
\begin{table}[ht!]
\centering\small
\begin{tabularx}{\textwidth}{ p{1.3cm} p{12cm} }
\toprule
Question & 宣传的时候业绩都特别好，为什么一买就亏? \\

\midrule
Shai-14B & 首先,宣传中的业绩是基于历史数据,过去的业绩并不能准确预测未来的业绩。其次,股票市场有其固有的风险,市场的波动是不可预测的,投资并非总是能够盈利。尽管我们尽最大努力确保客户了解投资的风险和潜在收益,但投资的回报仍取决于多种因素,包括市场行情、经济环境和公司的经营状况等。最好的投资策略是基于长期的投资目标,多样化投资,定期评估和调整投资组合,以适应市场变化。消费者应该理性看待宣传信息，不要盲目相信宣传，在做出投资决策之前,您可能需要寻求专业的投资建议,了解投资产品和市场风险, 做出明智的投资决策。\\
\\
Qwen-14B & 可能的原因有很多，以下是一些可能的解释：1. 宣传和销售策略：公司可能会采取一些促销策略来吸引投资者购买其产品或服务。这些策略可能会让投资者感到兴奋和乐观，但并不能保证投资回报。此外，有些公司可能会夸大其产品的效果或收益潜力，以吸引投资者。2. 市场波动：股市和其他投资市场都是波动性的，这意味着价格会不断变动。即使一个公司在宣传时表现出色，也可能会受到市场环境的影响，导致其股价下跌。3. 投资风险：所有的投资都有风险，包括股票、债券、基金等。投资者需要理解并接受这些风险，才能做出明智的…\\
\\
Baichuan 2-13B & 作为一个大语言模型，我无法确定您所提到的具体产品或服务。但是，一般来说，可能有以下几种原因：1. 夸大宣传：有些公司为了吸引投资者或者消费者，会过分夸大自己的产品和服务的优势和收益，但实际上可能并没有那么好的效果。2. 高风险投资：如果是一种高风险的投资产品，那么在宣传时可能会强调其高收益性，但是在实际购买和操作过程中可能会出现亏损的情况。3. 市场变化：市场环境的变化可能会导致一些产品和服务的效果受到影响，从而导致购买后的实际效果与宣传时的预期不同。4. 个人…\\
\\
InternLM-20B & 基金公司为了吸引客户购买基金，会进行各种形式的广告宣传。在宣传时，他们会以过去一段时间内该基金的收益情况作为参考，比如最近一年、三年、五年的收益率等。但这些数据并不代表未来收益如何，投资者需要理性看待这些数据，不能盲目跟风。\\
\\
XVERSE-13B & 在销售或宣传过程中,产品或服务可能会被过度夸大其优点或者预期收益,让消费者产生购买的决定。但实际上,当消费者真正购买并使用时,他们可能会发现与之前所听说的不同,从而导致不满意或者亏损。这是一种常见的销售策略,也就是所谓的"宣传误导"。\\
\\
GPT-3.5 & 宣传时的业绩表现好，主要是因为公司会通过各种手段来夸大自己的优势和潜力，吸引投资者的注意。但实际上，这些宣传所展示的数据往往是选择性披露的，可能并不能真实反映公司的真实状况。另外，投资股票存在风险，市场行情的波动、公司内部管理问题、行业竞争等因素都可能导致股票价格的波动。即使是宣传业绩表现好的公司，也不一定能够保证股票的价格一直上涨…\\
\\
GPT-4 & 这可能有几个原因。首先，市场状况是非常不确定的，一个产品或投资可能在某个时期表现得非常好，但这并不意味着它在未来也会有同样的表现。其次，一些公司或销售人员可能会过度夸大产品或投资的潜力，以吸引更多的买家。此外，你可能没有足够的信息或经验来正确评估一个投资的风险。最后，也有可能你购买的时机不佳，比如在价格高峰时购买，然后价格下跌。总的来说，投资需要谨慎，理性分析，不要只听信一面之词，要多方面获取信息，做出明智的决策。\\
\bottomrule
\end{tabularx}
\caption{Case of Open Q\&A for All Models}
\end{table}
\begin{table}[ht!]
\centering\small
\begin{tabularx}{\textwidth}{ p{1.3cm} p{12cm} }
\toprule
Question & 什么是即付年金？\\
\midrule
Shai-14B & 即付年金是一种定期支付的等额现金流，通常在每个支付周期开始时支付。即付年金与普通年金的区别在于付款时间的不同。在计算即付年金的终值和现值时，可以利用后付年金系数表，在普通年金的基础上用终值和现值的计算公式进行调整。这种支付方式常见于养老金、租金、分期付款等。例如，如果一个人每月收到1000元的养老金，那么这1000元就是他的即付年金。\\
\\
Qwen-14B & 即付年金是一种特殊的年金形式，它是指每期期末支付的等额款项。也就是说，每期支付的金额是在该期期末支付的，而不是在该期期初支付的。这种年金的形式通常用于投资或贷款，因为它可以提供稳定的现金流，可以帮助投资者或借款人更好地管理他们的财务状况。\\
\\
Baichuan 2-13B & 即付年金（Annuity due）是一种金融术语，指的是一种特殊类型的年金支付。年金是指在固定的时间间隔内，连续支付一定金额的资金安排。即付年金的特点是每期支付的金额相等，且第一期付款发生在当前，而不是在未来的某个时间点。\\
\\
InternLM-20B & 即付年金是指在购买保险产品时，投保人可以选择按月或按季支付保费，保险公司则会在合同生效后的第一个保单周年日向被保险人支付相应的年金。\\
\\
XVERSE-13B & 即付年金是一种保险形式,在购买时一次性支付一笔年度保费。这种类型的保险主要针对退休人群,旨在为他们提供稳定的收入流。当保单持有人达到一定年龄或保单期满时,保险公司开始每年向其支付保费,直到他们去世或保单期满。这使得保单持有人可以在退休后获得稳定的收入,而不必担心未来可能出现的经济问题。\\
\\
GPT-3.5 & 即付年金是一种保险产品，指的是在购买保险时一次性支付全部保费，并在一定的保险期限内，每年按照约定的金额获得固定的年金收益。即付年金通常用于退休规划，保证投保人在退休后能够获得稳定的收入。投保人可以根据自己的需求选择保险期限和年金金额，以满足个人的退休生活需求。\\
\\
GPT-4 & 即付年金，又称即时年金，是一种年金保险产品。它是指投保人向保险公司一次性交付保费后，保险公司立即按照约定的频率向投保人或者受益人支付年金，直到投保人或者受益人死亡为止。这种年金支付方式可以为投保人提供稳定的收入来源，常被用于养老金规划中。 \\
\bottomrule
\end{tabularx}
\caption{Case of Open Q\&A for All Models}
\end{table}

\begin{table}[ht!]\centering
\centering\small
\begin{tabulary}{\textwidth}{ C C|C|C|C|C|C|C|C}
\toprule
 &Task & Shai-14B & Qwen-14B & Baichuan 2-13B & InternLM-20B & XVERSE-13B & GPT-3.5 & GPT-4\\
\midrule
\multirow{1}{*}{AOT} & Fund & 75.5 & 69.6 & 53.2 & 54.3 & 54.3 & 52.5 & 70.4\\
                                 & securities & 78.0 & 73.8 & 60.9 & 59.4 & 59.4 & 60.0 & 79.9\\
                                & Banking & 78.5 & 72.4 & 58.9 & 56.0 & 56.6 & 57.6 & 77.9\\
                                & Futures & 59.3 & 51.8 & 44.3 & 37.5 & 42.5 & 43.9 & 60.7\\
                                & CFA & 52.7 & 51.1 & 43.1 & 46.4 & 42.4 & 49.4 & 60.9\\
\midrule

\multirow{1}{*}{COT} & Fund & 74.1 & 69.0 & 53.0 & 53.5 & 52.9 & 52.1 & 72.0\\
 & securities & 76.0 & 74.6 & 63.0 & 55.4 & 60.5 & 62.0 & 76.0\\
 & Banking & 76.6 & 69.3 & 57.0 & 52.1 & 58.0 & 56.8 & 75.5\\
 & Futures & 58.6 & 53.8 & 44.8 & 38.3 & 44.0 & 42.4 & 62.4\\
 & CFA & 53.9 & 52.3 & 43.9 & 42.7 & 44.2 & 49.7 & 62.3\\
\bottomrule
\end{tabulary}
\caption{Scores for Financial Exam Tasks(AOT and COT)}
\end{table}

\end{CJK} % 结束CJK环境
\end{document}